\documentclass[aps,pra,amsmath,amssymb,amsfonts,lengthcheck,twocolumn,superscriptaddress]{revtex4-1}
\usepackage{graphicx}
\usepackage{amsmath, amsthm, amssymb, amsfonts}
\usepackage{verbatim}
\usepackage{dcolumn}
\usepackage{bm}
\usepackage{epsf}
\usepackage{color}
\usepackage[colorlinks=true,citecolor=blue,linkcolor=blue,urlcolor=blue]{hyperref}%
\usepackage{xcolor}
\usepackage{physics}
\usepackage{dsfont}
\usepackage{tikz}
\usepackage{multirow}
\usepackage{xfrac}
\usepackage{makecell}
\usepackage{siunitx}
\usepackage{orcidlink}

\newcommand{\bla}{bla\\bla\\bla\\bla\\bla}

\DeclareMathAlphabet\mathbfcal{OMS}{cmsy}{b}{n}

\makeatletter
\newcommand{\currentfontsize}{\f@size pt}
\makeatother

\makeatletter
\newcommand\footnoteref[1]{\protected@xdef\@thefnmark{\ref{#1}}\@footnotemark}
\makeatother

\begin{document}

\title{Secure Quantum Handshakes: Quantum Network Verification Via Simon’s Algorithm}

\author{Reece Robertson\orcidlink{0000-0003-1064-0012}}
\affiliation{Department of Computer Science and Electrical Engineering, University of Maryland, Baltimore County, Baltimore, MD 21250, USA}
\affiliation{Department of Physics, University of Maryland, Baltimore County, Baltimore, MD 21250, USA}
\affiliation{Quantum Science Institute, University of Maryland, Baltimore County, Baltimore, MD 21250, USA}

\author{Sejal Patil}
\affiliation{Department of Information Science, University of Maryland, College Park, MD 20742, USA}

\author{Sebastian Deffner\orcidlink{0000-0003-0504-6932}}
\homepage{https://quthermo.umbc.edu/} 
\affiliation{Department of Physics, University of Maryland, Baltimore County, Baltimore, MD 21250, USA}
\affiliation{Quantum Science Institute, University of Maryland, Baltimore County, Baltimore, MD 21250, USA}
\affiliation{National Quantum Laboratory, College Park, MD 20740, USA}

\begin{abstract}
    Quantum networking represents the future of secure communication, enabling fundamentally new capabilities through quantum information exchange.
    However, these networks require secure, high-fidelity quantum communication channels.
    Currently, there is no reliable method to verify whether clients entering a quantum network truly possess quantum state communication capability over robust channels.
    In this work, we introduce a client-validation protocol that functions as a proof-of-quantum-communication handshake.
    Our protocol secures a quantum network against classical spoofers, faulty quantum computers, and genuine quantum devices connecting to the network over noisy or insecure channels.
    We validate the efficacy of this protocol through numerical simulations under erasure, dephasing, and depolarizing noise models.
\end{abstract}

\maketitle

\section{Introduction}

One of the most significant technological revolutions of the past century has been the advent of the modern computer network.
These networks have fundamentally changed the way people communicate, collaborate, do business, and experience the world.
Indeed, internet connection is often a prerequisite to participation in modern social spaces, from making a reservation to applying for a job to distributing an idea to a wide audience.

The second quantum revolution has changed the way we think about computers.
We have realized that information is physical \cite{Landauer1999}, and that our capacity to process information is limited only by the physical system used to represent it.
An emergent development in this revolution is the quantum network, which applies the same revolutionary ideas about quantum information processing to the methods by which we connect and share information.

The potential benefits of large, reliable quantum networks are myriad.
In the near-term, quantum networks may facilitate the deployment of large-scale important algorithms, such as Shor's factoring algorithm \cite{Shor1994}, across a collection of small-scale, fault-tolerant devices \cite{Lomonaco2004}.
Similar proposals exist for other significant problems, such as Simon's algorithm \cite{Zhou2025, Zhang2024}, and tools are being developed to compile algorithms for distributed systems \cite{Zhang2026FQAS}.
Additionally, a proposal exists to establish a quantum network with scientific utility but without cryptanalytic capacity \cite{Yadavalli2026}.
In the long term, quantum networks may speed up information transmission \cite{Harrow2004}, revolutionize the blockchain and other decentralized internet protocols \cite{Yang2024, Konara2026}, and enable a new quantum internet \cite{Wehner2018, Nath2026}.

Quantum communication, cryptography, sensing, and simulation likewise have military applications \cite{Krelina2021}, the moral and ethical dimensions of which are discussed, e.g., in Ref.~\cite{Derian2024}.
The use of quantum information networks for cryptographic communication dates back to as early as 1984 \cite{Bennett2014}, yet research on this problem remains ongoing today; see, e.g., Refs.~\cite{Sharma2021, Ghourab2025, Kao2026}.
Moreover, current research on the quantum cyber offensive includes an analysis of attacks on quantum repeater links \cite{Satoh2021}, distributed quantum key recovery attacks \cite{Zhang2026Distributed, Shang2026, Zhou2023}, and $\mathsf{SWAP}$-based attacks for interrupting the execution of quantum algorithms \cite{Lee2025}.

A prerequisite to any of these applications is the ability to construct reliable networks of quantum devices.
Consider the problem of constructing one such closed network comprised entirely of verified quantum computing nodes, each operating under mutually established trust.
Since admission to the network entails the trust and privileges afforded to all verified quantum members, incorrectly admitting a faulty node directly threatens the integrity of the network as a whole.
Thus, when an unknown external device requests admission, the network faces a fundamental verification problem: it must first verify that the device is a genuine quantum system and not a classical machine misrepresenting its capabilities, and it must second verify that the machine can communicate quantum information to the network members with sufficiently high fidelity.

The challenge, therefore, is not merely one of authentication in the conventional sense, but of physical verification---verification that cannot be determined through classical interrogation alone.
Any classical system can assert quantum capability through credentials, certificates, or protocol compliance.
What it cannot do is exhibit quantum behavior through the demonstration of statistical properties which are provably unreachable through classical means.
Likewise, a faulty quantum device, or a device communicating quantum information through a noisy or insecure channel, will similarly fail to demonstrate all statistical properties of fault-tolerant quantum systems.

The present paper presents a novel protocol for establishing a quantum network and verifying its integrity.
The protocol exploits the exponential separation between classical and fault-tolerant quantum systems in solving oracular quantum problems; this work demonstrates the protocol using Simon's period-finding algorithm \cite{Simon1997}.
Thus, rather than asking a candidate node to identify itself, the network instead requires it to solve a Simon problem using a number of oracle queries insufficient for a classical machine to solve the problem.
Admission is granted if and only if the node can successfully demonstrate \textit{quantum proclivity}---the ability to fault-tolerantly generate, transmit, and measure quantum states---through the identification of the period hidden within the oracle.
This ensures that it is effectively impossible for any classical node to gain network admittance.
Moreover, due to the complexity of the postprocessing routine for a noisy implementation of Simon's algorithm, a sufficiently sized oracle can also be used to reject genuine quantum nodes that fail to meet an error threshold---either due to noise on the node itself or noise in the communication channel between the node and the network.

This work builds upon existing literature in quantum networking.
A readable overview of the field is given by \citet{Bhuyan2024}.
An algebraic language for reasoning about quantum networks has been developed \cite{Buckley2024, Buckley2023}, and algorithmic tools exist for configuring entanglement distribution networks that minimize the cost incurred by users \cite{Gu2026}.
Moreover, a quantum network protocol designed for short-term quantum memory has been developed \cite{Kozlowski2020}.
These tools and techniques lay the foundation for reasoning about quantum networks, but they are not especially concerned with verifying the quantum capabilities of all nodes within the network.
Thus, a protocol to generate a verified quantum network fills an important omission that is prerequisite to these studies.

Additional work has been done on using classical computers to evaluate and verify quantum networks \cite{Gu2024, Mahadev2018}, or to model them using classical digital twins \cite{Elsokary2026}.
Another study uses quantum resources, via Grover's algorithm, for classical network verification \cite{Dozier2024}.
Our work here differs in that we use quantum resources to verify the integrity of quantum networks.

Finally, several proposals exist that apply quantum computation to the problem of quantum network verification; see \citet{Gheorghiu2021} for an overview of this work.
Significant to our purposes, multiple nonlocal game protocols exist which verify entanglement within quantum networks \cite{Luo2019, Cheng2021}.
Another protocol certifies the topology of quantum networks \cite{Weinbrenner2024}.
Perhaps most significantly, \citet{Alagic2023} use the term ``quantum network verification'' to refer to a protocol that allows two quantum devices to prove that they can share a quantum state.
Our work builds on this body of literature to contribute a protocol for the \textit{generation} of a verified quantum network that maintains high-fidelity communication between network nodes.

\section{Methods}\label{sec:methods}

With the relevant background established, we turn our attention to our protocol for quantum network verification.
We begin with a discussion of the challenge of network verification and our proposal to address it.
Following this, we discuss Simon's algorithm, which we choose as a model algorithm with which to demonstrate the protocol.
We then discuss an illustrative example of the protocol in action, and conclude with a series of additional considerations for implementation.

As with all elements of quantum computer science, the problem of network verification has an analog in classical computer science.
In the classical setting, the problem of network integrity is primarily expressed as the Byzantine generals problem of \citet{Lamport1982}, in which a collection of generals (network components) must decide on a plan of attack (a network action) in the presence of traitorous generals (faulty or malicious network nodes).
See Ref.~\cite{Ben2005} for a quantum version of the Byzantine generals problem. In addition, another challenge in classical network verification is the remote resource attestation problem, in which a state of a system must be verified remotely \cite{Banks2021}, which has also been generalized to a collection of quantum devices that verify shared quantum resources  \cite{Ganardi2026}.

Our protocol addresses a problem that is related to, but distinct from, both Byzantine agreement and remote resource attestation.
We are interested in verifying the fault-tolerant quantum proclivity of candidate nodes prior to incorporating them into a network.
This is different from the Byzantine generals problem, as we are discriminating who to include at the war table (in the Byzantine problem, the generals are known and their number is fixed throughout the process).
It is likewise different from remote resource attestation, as we are verifying quantum \textit{capability}, not quantum states, entanglement, coherence, or other resources.

\subsection{Quantum Network Verification}

An ideal quantum network consists of fault-tolerant quantum devices (or nodes) connected via secure, high-fidelity quantum channels.
Quantum network verification, then, is the problem of validating that a quantum network is composed of such fault-tolerant devices and connections.
The verified status of a network can be established inductively: a fault-tolerant quantum device can conditionally admit other nodes into the network only after rigorously verifying their proclivity to fault-tolerantly process and exchange quantum information.

Our novel protocol begins with a verified quantum device, hereafter called the quantum server.
The server accepts messages from various clients that request admittance to the network.
When the server receives a request, it instantiates an oracle for an oracular quantum algorithm and allows the client to query the oracle over a quantum communication channel.
If the client can successfully recover a piece of information hidden within the oracle which is inaccessible through classical interrogation, then the server admits the client to the network, having verified the fault-tolerant quantum proclivity of the client and the robustness of the quantum communication channel.
If the client cannot recover the information, then the server refuses to admit the client to the network.

\begin{figure}[t]
    \centering
    \includegraphics[width=\linewidth]{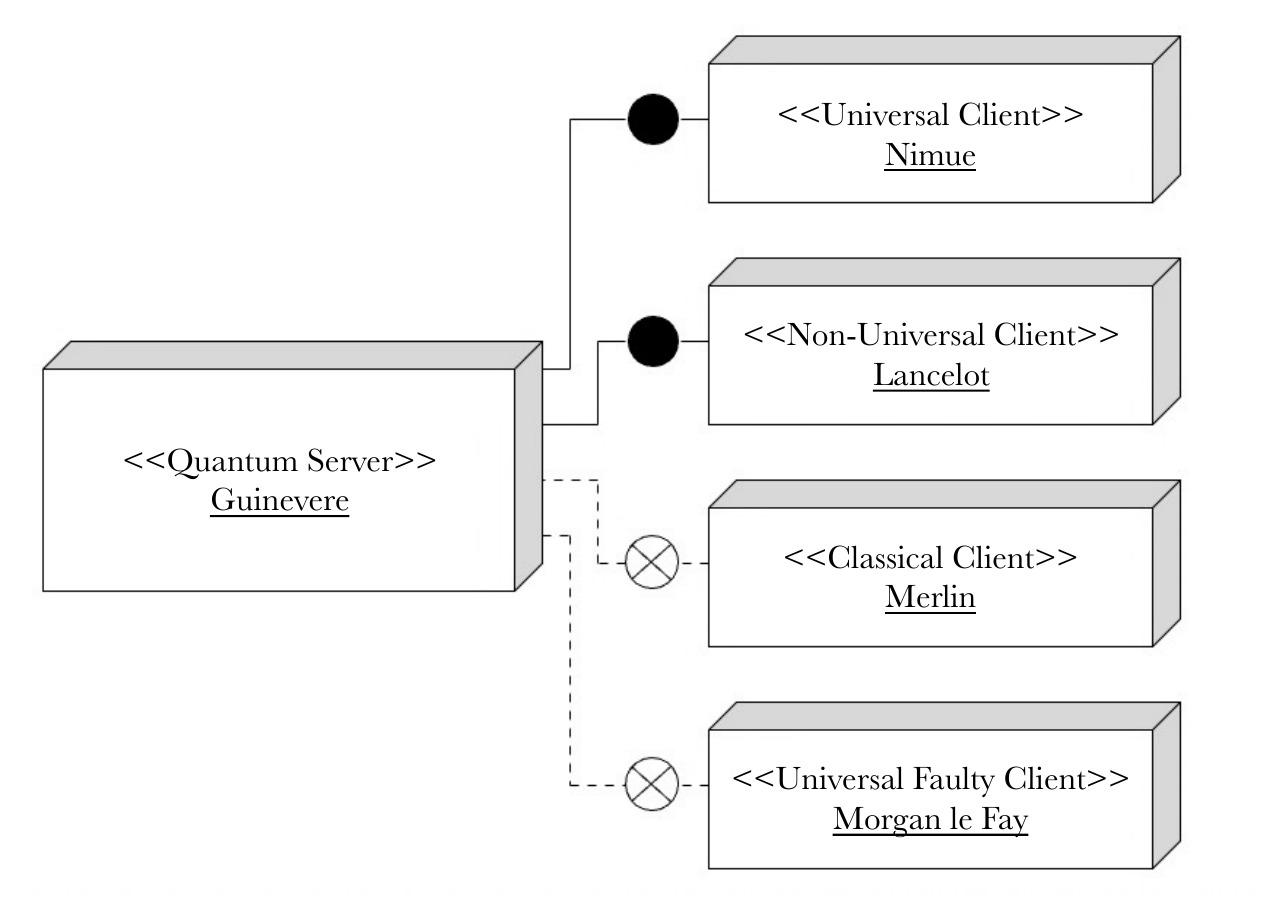}
    \caption{
        \textbf{Verified quantum network.}
        An example of a verified quantum network constructed via our protocol.
        Here Guinevere represents a quantum server, Nimue a universal and fault-tolerant quantum client, Lancelot a non-universal yet fault-tolerant quantum client, Merlin a classical client, and Morgan a universal and faulty quantum client.
        Through the protocol, Guinevere validates Nimue's quantum capacity and admits her into the network.
        By the same token, Guinevere rejects Merlin and Morgan from the network when they fail to show fault-tolerant quantum capabilities.
    }
    \label{fig:network-example}
\end{figure}

In the end, this protocol allows for the creation of a verified network of fault-tolerant quantum devices, while both classical and faulty quantum devices are eliminated, as shown in Figure \ref{fig:network-example}.
In Section \ref{sec:example}, we explore the interaction of the server with each of these potential network clients.
For clarity of presentation, in the remainder of this paper we identify characters from the Matter of Britain with each of the candidate network components.
We choose to use the name of Queen Guinevere to denote the universal, fault-tolerant quantum server.
Likewise, we use the names of four magical beings---Nimue, Lancelot, Merlin, and Morgan---to denote each of four potential clients.
The capabilities of each client are enumerated below.
\begin{itemize}
    \item Nimue, or the Lady of the Lake, is a famous fairy of the King Arthur legend.
    Thus, we choose to use her name to denote a universal, fault-tolerant quantum computing machine which is capable of communicating with Guinevere over a secure, fault-tolerant, entanglement-preserving channel.
    \item Lancelot du Lac, a preeminent knight of the Round Table, is (in some manuscripts) raised in the fairy realm as Nimue's adopted son.
    As is fitting to this relationship, we use Lancelot's name to denote a quantum computer which is fault-tolerant, but not necessarily universal.
    In particular, a Lancelot device needs only possess the capacity to locally and fault-tolerantly prepare and measure $X$-basis states.
    Lancelot must also be capable of communicating quantum states to Guinevere over a fault-tolerant, but perhaps entanglement-breaking, quantum channel.
    Lancelot need not be able to perform universal quantum computation, nor need he be able to create or preserve entanglement.
    In this paper, we use the term \textit{quantum proclivity} to denote a device classified as either a Nimue or a Lancelot.
    \item Merlin, the famous magician, is frequently described as a mad prophet.
    In this work, we use his name to denote a classical computer which falsely claims quantum proclivity.
    In particular, we assume that Merlin can communicate with Guinevere over a secure quantum channel, but, significantly, Merlin can only send and receive classical information over this channel.
    \item Morgan le Fay is often described as a fairy enchantress who is an enemy to Arthur.
    We use her name to denote a faulty quantum machine which is otherwise capable of universal quantum computation.
    Morgan's faults may be local, or due to errors incurred during communication with Guinevere, perhaps due to eavesdroppers in that channel.
\end{itemize}

Finally, any oracular quantum algorithm may serve as the basis with which the server discriminates quantum clients from classical clients.
In this paper, we simulate our proposed protocol using Simon's algorithm as the oracle function.
Simon's algorithm is desirable for this role for several reasons: it is easy for Guinevere to implement, Nimue and Lancelot can crack the oracle with a linear number of queries and a polynomial amount of classical post-processing, and there is little motivation to dequantize this algorithm.
We review the fundamentals of Simon's algorithm in the next subsection.

\subsection{Simon's Algorithm}

\begin{figure*}[t]
    \centering
    \includegraphics[width=0.7\linewidth]{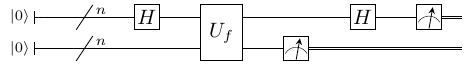}
    \caption{
        \textbf{Simon's algorithm.}
        The circuit diagram for Simon's algorithm.
        This circuit involves two $n$-qubit quantum registers collocated on a single quantum processor.
        The algorithm uses transversal measurements and Hadamard transformations.
        It also assumes access to a black-box unitary function $U_f\left(|x_1\rangle|x_2\rangle\right) = |x_1\rangle|x_2\oplus f(x_1)\rangle$ which implements the unknown oracle of \eqref{eq:oracle} and \eqref{eq:constraint} as a reversible unitary operation for execution on a quantum device.
        At the conclusion of a fault-tolerant implementation of the algorithm, the bitstring measured in the top register will be a sample from \eqref{eq:orthogonal-set}.
        Both this and the following circuit diagrams were created using $\langle\text{q}|\text{pic}\rangle$ \cite{Draper2016}.
    }
    \label{fig:simons}
\end{figure*}

\begin{figure*}[t]
    \centering
    \includegraphics[width=0.95\linewidth]{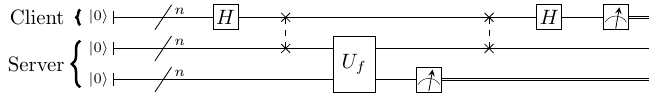}
    \caption{
        \textbf{Networked Simon's algorithm.}
        A networked implementation of Simon's algorithm; compare to the original version presented in Figure \ref{fig:simons}.
        In this version, three $n$-qubit quantum registers are used, one of which is controlled by a client device while the other two are controlled by a quantum server.
        The client prepares a superposition state which it sends to the server through a quantum channel, as denoted by the dashed $\mathsf{SWAP}$ gates.
        The server computes the oracle on the input and sends the result back to the client over the communication channel.
        The client may then prepare and sample from \eqref{eq:orthogonal-set} as normal.
        This enables a situation in which the Simon oracle is implementable on the server yet remains entirely unknown to the client, allowing the server to probe the quantum capacity of the client device.
    }
    \label{fig:networked-simons}
\end{figure*}

Simon’s algorithm provides a polynomial-time solution to Simon's problem, a hidden-subgroup problem for which the fastest known classical solution runs in exponential time \cite{Simon1997}.
Simon's problem begins with a two-to-one black-box oracle function
\begin{equation}
    f:\{0,1\}^n\rightarrow\{0,1\}^n
    \label{eq:oracle}
\end{equation}
that acts on bitstrings of size $n$.
We are promised that there exists a secret string $s\in\{0,1\}^n$ such that for any two distinct $n$-bit strings $x,x'\in\{0,1\}^n$, it follows that
\begin{equation}
    f(x)=f(x') \iff x' = x \oplus s,
    \label{eq:constraint}
\end{equation}
where $\oplus$ denotes the bitwise $\mathsf{XOR}$ function.
The challenge of Simon's problem is to identify $s$ when $f$ is entirely unknown and can only be queried.

Classically, recovering $s$ in this setting requires a number of oracle queries that scales exponentially in $n$ \cite{Simon1997}.
However, given access to a quantum computer and an unknown yet implementable unitary function $U_f$, defined such that
\begin{equation}
    U_f(|\psi_1\rangle|\psi_2\rangle) = |\psi_1\rangle|\psi_2\oplus f(\psi_1)\rangle,
\end{equation}
one can use Simon’s algorithm to solve the problem with a linear number of oracle queries.

Simon's algorithm achieves this exponential speedup by using the oracle queries to prepare and sample from the set
\begin{equation}
    \left\{\omega\in\{0,1\}^n\textbf{ }\Bigg|\textbf{ }\omega\cdot s = \left(\sum_{i=1}^n \omega_is_i\right)\mod 2 = 0 \right\}.
    \label{eq:orthogonal-set}
\end{equation}
Given $n-1$ nontrivial independent samples of \eqref{eq:orthogonal-set}, one can construct a linear system of binary equations that, once solved, reveals the hidden string $s$ \cite{Simon1997}.
The quantum routine uses a number of operations that scales polynomially with $n$, and the classical linear system post-processing routine scales at worst cubically in $n$ \cite{Nielsen2010}.
Additionally, in Appendix G of his textbook, \citet{Mermin2007} proves that with $n+20$ shots of (a fault-tolerant implementation of) Simon's algorithm, the probability of failing to obtain $n-1$ nontrivial independent samples of \eqref{eq:orthogonal-set} is less than one in a million, regardless of the size of $n$.
Therefore, the entire complexity of Simon's algorithm is polynomial in $n$, which accounts for the exponential speedup over the classical method.

The circuit diagram that implements Simon's algorithm on a single quantum processor is given in Figure \ref{fig:simons}.
This diagram also reveals a fundamental tension in Simon's algorithm: the oracle must be simultaneously unknown and implementable.
As explained by \citet{Aaronson2018}, there is no known instantiation of an oracle satisfying the constraints of Simon's problem that does not immediately admit a faster classical solution to the problem.
As such, the quantum advantage is predicated on the assertion that the implementation of the oracle is unknown.
However, a truly unknown oracle renders a local implementation of the oracle impossible.

This tension is resolved when we consider a networked implementation of Simon's algorithm.
A circuit diagram for such an implementation is given in Figure \ref{fig:networked-simons}.
The dotted $\mathsf{SWAP}$ gates used in the circuit denote transmission of an $n$-qubit quantum state over a quantum channel.
Notice that the algorithm is logically equivalent to the original implementation of Simon's algorithm shown in Figure \ref{fig:simons}; the difference between these circuits is the introduction of client and server components of the algorithm, where the computation of the oracle is localized on the server.
This allows for the client to query the oracle without knowledge of the implementation details of the oracle---a feature which resolves the paradox of Simon's algorithm while simultaneously allowing the server to probe the quantum capacity of the client.

\subsection{Illustrative Example}\label{sec:example}

We now consider the execution of our proposed protocol for network verification from the perspective of Nimue, Lancelot, Merlin, and Morgan in turn.
In the discussion that follows, we use the symbol $x$ when referring to an ensemble of $n$-bit strings, and we use the symbol $z$ when referring to a single fixed $n$-bit string.

\subsubsection{Nimue \& Lancelot: Fault-Tolerant Quantum Clients}

\begin{figure}[t]
    \centering
    \includegraphics[width=\linewidth]{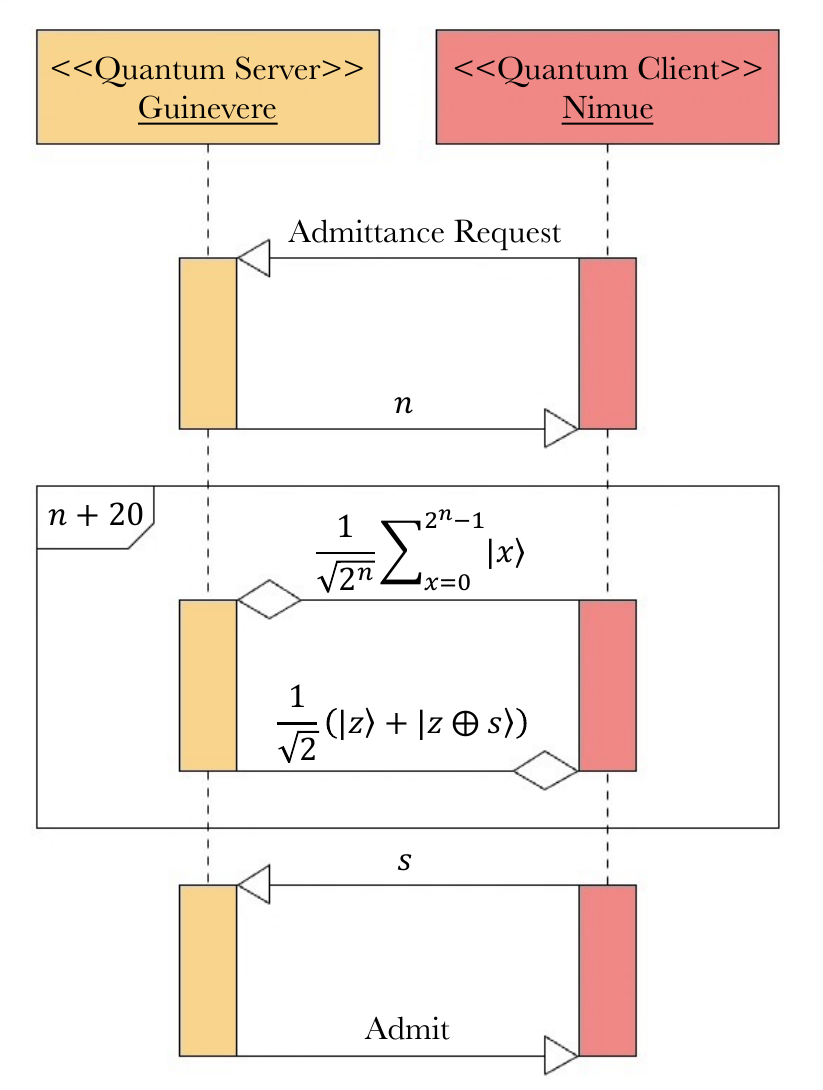}
    \caption{
        \textbf{Network verification protocol.}
        The communication timeline diagram between Guinevere, the quantum server, and Nimue, the universal, fault-tolerant quantum client.
        Note that Lancelot, who is not universal yet possesses quantum proclivity, proceeds through this protocol identically to Nimue.
        The triangular arrows represent classical messages, and the diamond arrows represent quantum messages.
        The messages inside the box are repeated $n+20$ times.
        Through this protocol, Guinevere is able to verify the quantum proclivity of Nimue and Lancelot, and the lack of quantum proclivity in Merlin and Morgan.
    }
    \label{fig:ideal-protocol}
\end{figure}

Nimue represents a universal, fault-tolerant quantum device communicating with Guinevere over a secure quantum channel.
Lancelot represents a non-universal quantum client which demonstrates quantum proclivity as it has been defined above.
Both clients successfully pass the protocol and are admitted to the network in the same way.
The interactions of these clients with Guinevere throughout the protocol are described below and summarized in the communication timeline diagram of Figure \ref{fig:ideal-protocol}.
For clarity of presentation, going forward we present the protocol from the perspective of Nimue.

First, Nimue uses a (public) classical communication channel to ask Guinevere for admission into the quantum network.
Guinevere chooses an $n$, $f$, and $U_f$ with which to probe Nimue's quantum capability, and she responds to Nimue's initial message with $n$.

Nimue then prepares a collection of $n$-qubits in an even superposition over all bitstrings of size $n$,
\begin{equation}
    \frac{1}{\sqrt{2^n}}\sum_{x=0}^{2^n-1}|x\rangle,
    \label{eq:even-superposition}
\end{equation}
which she sends to Guinevere over the quantum channel.
After receiving Nimue's query, Guinevere computes
\begin{equation}
    U_f\left(\frac{1}{\sqrt{2^n}}\sum_{x=0}^{2^n-1}|x\rangle|0\rangle\right) = \frac{1}{\sqrt{2^n}}\sum_{x=0}^{2^n-1}|x\rangle|f(x)\rangle.
\end{equation}
Guinevere measures the second register to select one random $|f(z)\rangle$, which, by entanglement, causes her first register to collapse to the state 
\begin{equation}
    \frac{1}{\sqrt{2}}\left(|z\rangle + |z\oplus s\rangle\right).
    \label{eq:post-measure}
\end{equation}
Guinevere returns \eqref{eq:post-measure} over the quantum channel, and Nimue uses it to draw a sample from \eqref{eq:orthogonal-set}.

Guinevere and Nimue repeat this quantum information transaction $n+20$ times.
Following this, Guinevere refuses to provide Nimue with any further oracle evaluations, and demands that Nimue send her $s$.
As mentioned above, with a probability of failure less than one in a million, Nimue has obtained sufficient information to reconstruct $s$ and return it to Guinevere over a classical channel.
Once Guinevere has validated Nimue's reconstructed $s$, she admits Nimue into the network as a verified quantum device.
Moreover, Guinevere immediately discards $f$ and $U_f$, so as to defeat any Merlins or Morgans who may be eavesdropping undetected over the classical channel.

\subsubsection{Merlin: Classical Client}

We now consider this protocol from the perspective of Merlin, who represents a classical computer falsely claiming quantum capacity.
Merlin can connect to Guinevere via a quantum communication channel, but he can only send classical information over this channel.

Since Merlin has only classical resources, he cannot prepare \eqref{eq:even-superposition}.
Instead, he sends Guinevere the state $|z\rangle$ for some $z\in\{0,1\}^n$.
Guinevere computes
\begin{equation}
    U_f(|z\rangle|0\rangle) = |z\rangle|f(z)\rangle,
\end{equation}
and, because measuring the second register does not affect the first register, she returns $|z\rangle$ to Merlin unchanged.
Therefore, Merlin has gained no information about the oracle $f$ whatsoever.
Any ``reconstructed $s$'' that Merlin sends over the classical channel is chosen with no information about the true $s$, hence Merlin can do no better than a random guess from the set of all nontrivial $n$-bit strings.

The probability that Merlin is able to guess $s$ at random can be made arbitrarily small through the appropriate choice of $n$.
With an oracle of size $n=20$, for example, the total number of possible bitstrings Merlin has to choose from is $2^{20}-1=1,048,575$, which gives him a success probability of less than one in a million.

\subsubsection{Morgan le Fay: Faulty Quantum Client}

Finally, Morgan executes the protocol in the same way as Nimue, with one significant difference---the noise in Morgan's computations prevents her from sampling from \eqref{eq:orthogonal-set} with certainty.
The initial state that she sends to Guinevere will have some error, and the response she receives will be only approximately \eqref{eq:post-measure}.
This means that the state Morgan samples with her oracle query may contain elements outside of \eqref{eq:orthogonal-set} that have an appreciable chance of being measured \cite{Robertson2025NISQ}.

As the probability that Morgan samples an invalid state grows, the likelihood that she fails to recover the proper value of $s$ also grows \cite{Robertson2025NISQ}.
The simulations shown in Section \ref{sec:simulations} establish that (given an assumption about Morgan's noise model) Guinevere can select $n$ such that Morgan will succeed only if she can perform quantum computations above an arbitrarily high fidelity threshold.
Guinevere then conditionally admits Morgan to the network if she passes the test.

As a final note, observe that the noise in Morgan's communication can arise from her device, from her communication channel, or from an eavesdropper in her quantum communication.
It makes no difference to Guinevere where the noise arises.
The result, however, is the same; Guinevere cannot trust Morgan's computations as they are inaccurate and insecure.

\subsection{Additional Considerations}

We conclude this discussion with a few remarks about possible modifications to our protocol.

First, in all of the prior discussion, Guinevere challenged each candidate client with exactly one oracle.
Guinevere could instead challenge candidates with multiple distinct oracles, which allows her to tighten security without increasing $n$.
For example, suppose that a Nimue (or Lancelot), Merlin, and Morgan each request to join the network, and Guinevere generates oracles with $n=15$ to validate them.
Assume further that Morgan has a $67\%$ probability of recovering $s$ for any oracle.
If Guinevere requires each client to pass one oracle, then Nimue will succeed with a probability greater than $99.9999\%$, Merlin will succeed with a probability less than $0.0031\%$, and Morgan will succeed with a probability of $67\%$.
If instead Guinevere requires 18 successes, Morgan will succeed with a probability less than 1\%, Merlin will fail with certainty, and Nimue will succeed with a probability greater than $99.9982\%$.
However, with 18 successes required, a Morgan with a $90\%$ chance of success for each oracle has a $15\%$ probability of being admitted to the network, and a Morgan with a $99\%$ chance of success for each oracle has over an $83\%$ chance of admission.
Consequently, Guinevere can control the quality of admitted quantum devices through the number of oracle challenges.

Second, we have so far considered networks with a single server and some number of clients.
A more robust version of this protocol requires each network node to play the role of Guinevere for each client requesting network admission.
Under this formula, every admitted client demonstrates capacity for high-fidelity quantum information exchange with all network nodes.
Of course, we need not insist that \textit{all} network nodes validate new clients; we could instead require each candidate client to successfully pass an oracle challenge with the nodes ``closest'' to it.
This allows for some flexibility in the network topology while maintaining a set of verified communication channels with which to distribute quantum information throughout the network.

Finally, this entire protocol assumes access to a network capable of transmitting quantum states.
While such networks have been demonstrated (e.g., \cite{Yin2017, Wu2026, Ferrari2026}), widespread public access to such networks has not yet been realized.
For this reason, we also implement a variant of the algorithm which can be performed entirely over classical communication channels.
This alternate protocol is outlined in Appendix \ref{sec:approximate-protocol}.

\section{Simulations}\label{sec:simulations}

We conclude the analysis of the proposed protocol by demonstrating its viability through numerical simulation.
It would be ideal to validate the protocol through real quantum network connections; however, this is not possible with the resources available in the current public quantum landscape.

We simulate the protocol with a faulty quantum client (a Morgan) subject to three distinct sources of noise: an erasure model, a dephasing model, and a depolarizing model.
These models are chosen because they are representative of the noise sources for photonic quantum communication \cite{Aifer2023, Meystre2021}, which encompasses virtually all quantum communication.
We report on the probability of successfully recovering the secret string $s$ of a Simon oracle as a function of the noise parameter $p$.

For each model, we sweep $p$ over the range $[0,0.1]$, and we sweep $n$ over $[2,15]$.
For each combination of $n$ and $p$, we instantiate 100 Simon's oracle problems, and draw $n+20$ samples of \eqref{eq:orthogonal-set} under the relevant noise model.
We then attempt to solve each noisy system by subsampling a set of $n-1$ equations which we try to solve exactly.
If a nontrivial solution to the subsystem exists, then we ``submit'' this as our reconstructed $s$; if it is correct, then this represents a success for Morgan; otherwise this is a failure.
Moreover, if 50 independent subsamples of the noisy system are unable to yield a system of equations with a nontrivial solution, then we also consider this to be a failure for Morgan.
The probability of recovering $s$ in this way for each noise model is given in Figures \ref{fig:photon-loss}--\ref{fig:depolarizing}.

\subsection{Erasure Model}

First, we consider an erasure model.
This model is perhaps the most relevant of the three models considered, as it is likely that this protocol will be deployed on quantum networks where a dominant source of error is atmospheric turbulence in free-space quantum communication, as has been shown in many studies \cite{Heim2010, Bohmann2017, Yuan2018, Vankov2020, Lanning2021, Xu2021, Ghalaii2022, Mao2022, Wilson2022, Klen2023, Chakraborty2024, Waghmare2025, Pechonkin2025}.

To model this type of noise, for each $z$ sampled from \eqref{eq:orthogonal-set}, we corrupt $z$ in the following way.
For each bit $z_i$ of $z$, we draw a $\pi_i$ from the uniform distribution on $[0,1]$.
We then conditionally erase $z_i$ if $\pi_i<0.5$, that is,
\begin{equation}
    z_i = \begin{cases}
        z_i&\text{ if }\text{ }\text{ }\pi_i > p, \\
        \emptyset&\text{ if }\text{ }\text{ }\pi_i \leq p,
    \end{cases}
\end{equation}
where $\emptyset$ denotes a null value that is neither 0 nor 1.
Note that this abstracts photon loss, which could occur in both outgoing and incoming messages, into a single event applied to the client's incoming message (after measurement in a classical basis).
Finally, note also that when Morgan is solving a subsystem that includes an equation with missing information, she fills the information with a random choice of 0 or 1.

\begin{figure}
    \centering
    \includegraphics[width=\linewidth]{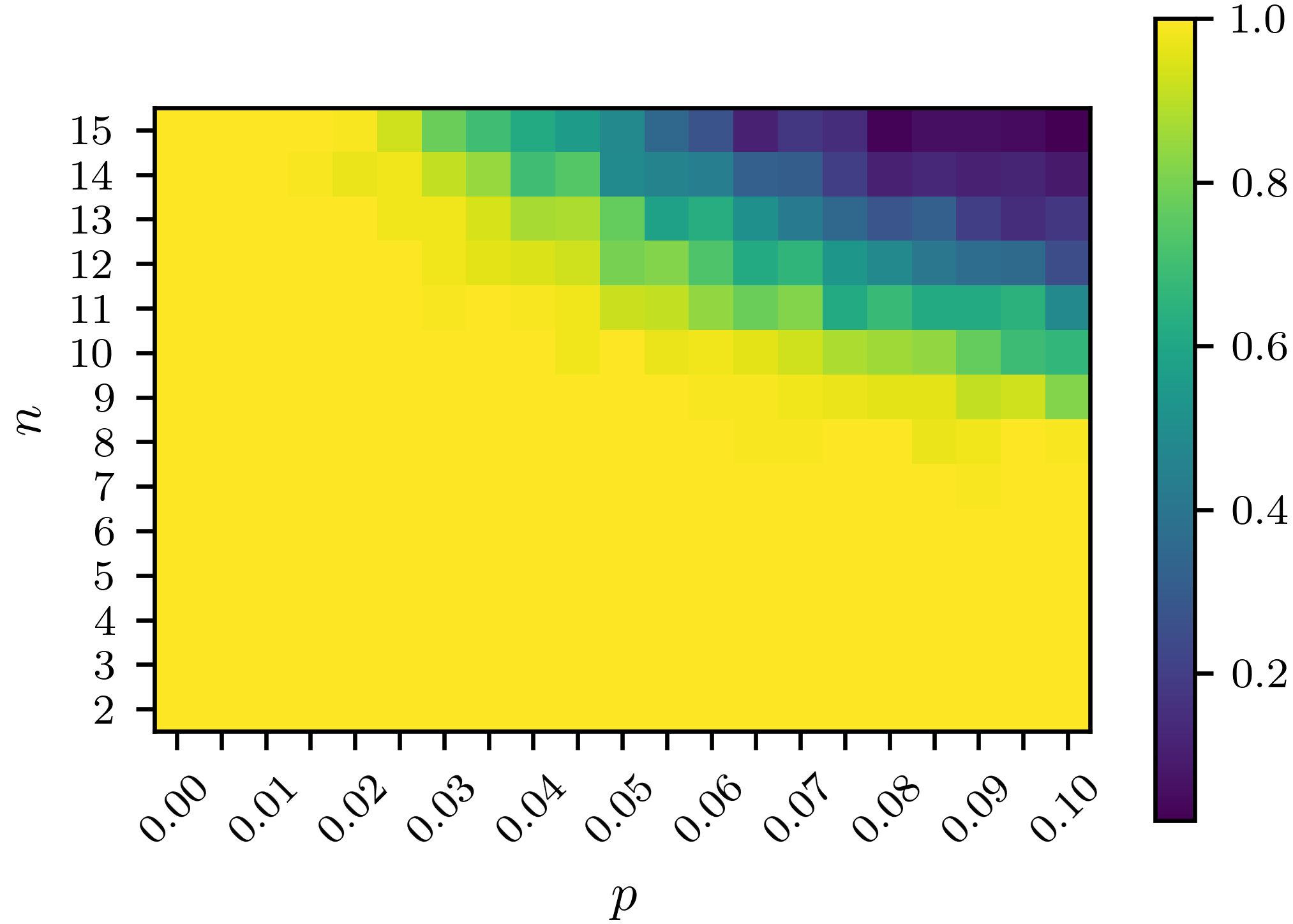}
    \caption{
        \textbf{Admittance probability under an erasure channel.}
        The results of a simulation of this protocol under a classical erasure noise model, which is typical of quantum communication via photons.
        The solution probability is plotted as a function of the problem size $n$ over the strength of the noise channel $p$.
        As expected, we see ideal performance in the noiseless case, and we see the probability of success decay as both $p$ and $n$ increase.
    }
    \label{fig:photon-loss}
\end{figure}

Morgan's successful reconstruction of $s$ under this noise model is shown in Figure \ref{fig:photon-loss}.
Observe that in the case where $p=0$, Morgan (or rather Nimue) succeeds with certainty.
This is as expected; when noise is absent, the probability of failure is less than one in a million.
That said, as $p$ increases, and also as $n$ increases, we see the probability of success decrease.
Again, this is as expected; identifying the solution becomes more difficult as the system of equations becomes more noisy.
Finally, as we will see, the probability of success under this source of noise is on average higher than the probability of success under the other noise sources.
This is likely due to the fact that under this model the locations of the errors are known, which enables more efficient decoding.

\subsection{Dephasing Model}

As a second noise model, we consider a classical dephasing model applied to the readout.
That is, for each $z$ sampled from \eqref{eq:orthogonal-set}, we corrupt $z$ a bit at a time as follows:
\begin{equation}
    z_i = \begin{cases}
        z_i&\text{ if }\text{ }\text{ }\pi_i > p, \\
        1 - z_i&\text{ if }\text{ }\text{ }\pi_i \leq p.
    \end{cases}
\end{equation}
Notice that this is a bit-flip model in the computational basis, which is a dephasing model in the Bell basis.
For this model, all quantum simulation and communication are assumed to be ideal---the bit flips, with an independent probability $p$ for each flip, are the only source of error.
This model is in spirit very similar to the photon loss model, except that here the location of each error is unknown.
As such, there is a general decrease in success probability across the board.

\begin{figure}
    \centering
    \includegraphics[width=\linewidth]{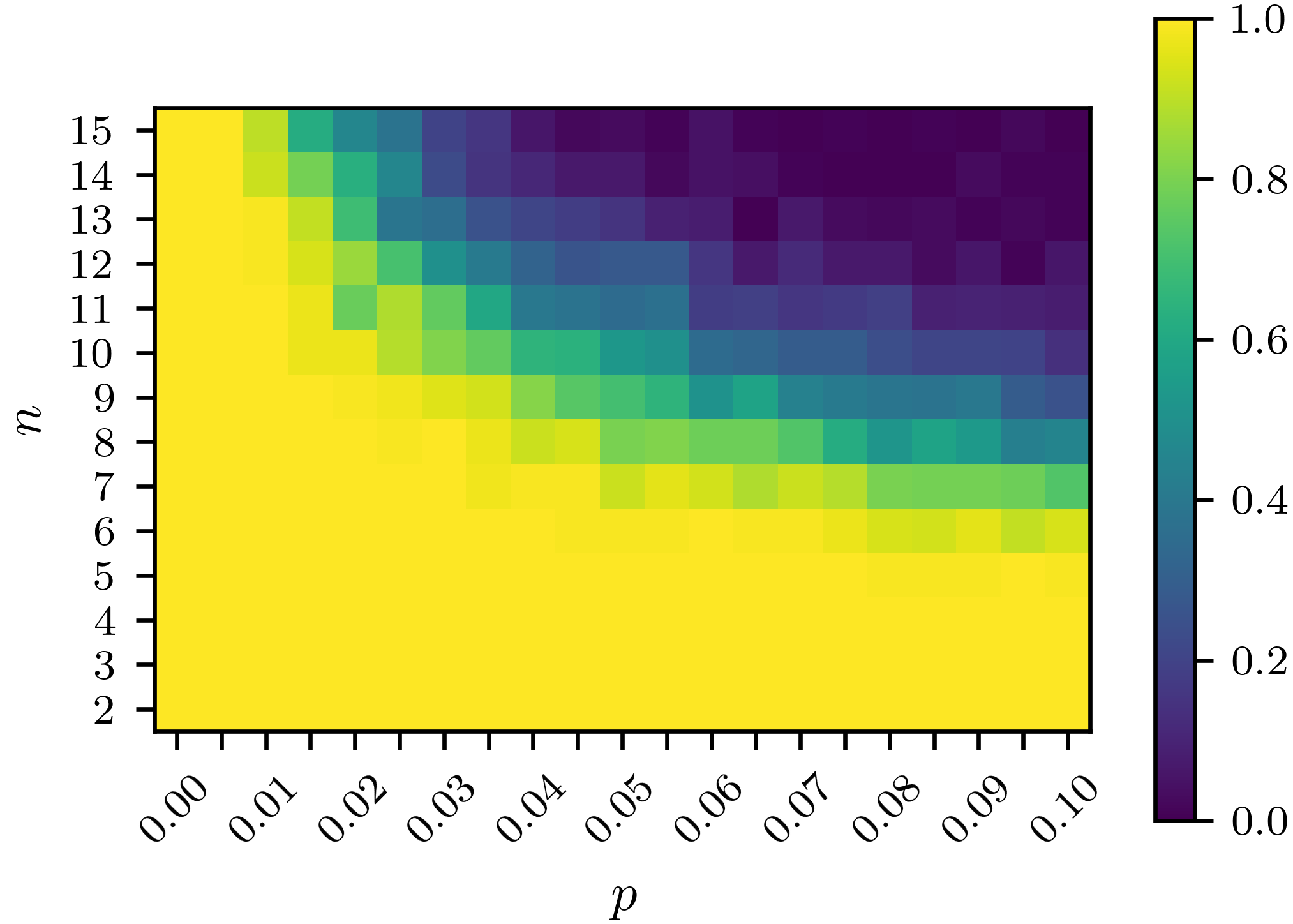}
    \caption{
        \textbf{Admittance probability under a dephasing channel.}
        The results of a simulation of this protocol under a classical dephasing (bit-flip) noise model, with solution probability plotted as a function of the problem size $n$ over the strength of the noise channel $p$.
        As before, we see ideal performance when $p=0$, and a decay in performance as $p$ and $n$ increase.
    }
    \label{fig:bit-flip}
\end{figure}

The results are given in Figure \ref{fig:bit-flip}.
Here again, we see that $p=0$ successfully recovers fault-tolerant quantum computation, as expected.
Additionally, we again see our performance degrade as both $n$ and $p$ increase.

\subsection{Depolarizing Model}

Finally, we consider a depolarizing model applied to all the one- and two-qubit gates in a simulation of Simon's algorithm.
In contrast to the prior two models, this is a truly quantum noise model.
Moreover, it models a situation where Morgan possesses a faulty quantum device which is able to communicate with Guinevere through a noiseless channel.
For ease of modeling, we applied a noise model to all the quantum gates, including those that implement the oracle.
While we in general assume that Guinevere has access to fault-tolerant quantum resources, this simulation is sufficient to probe the general character of this type of quantum noise in the protocol.

\begin{figure}
    \centering
    \includegraphics[width=\linewidth]{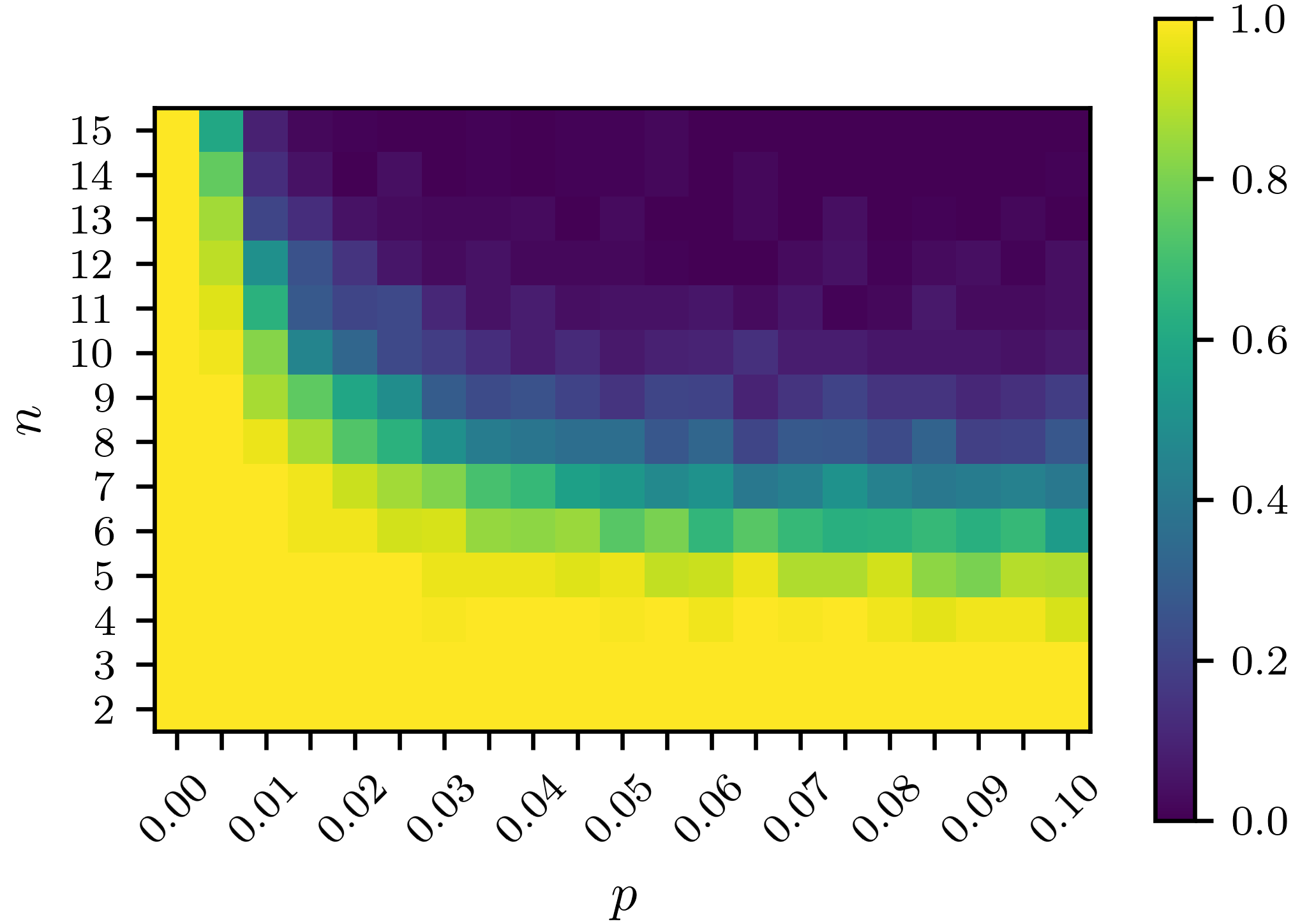}
    \caption{
        \textbf{Admittance probability under a depolarizing channel.}
        The results of a simulation of this protocol under a quantum depolarizing noise model, with solution probability plotted as a function of the problem size $n$ over the strength of the noise channel $p$.
        Again, we see ideal performance when $p=0$, and a decay in performance as $p$ and $n$ increase.
    }
    \label{fig:depolarizing}
\end{figure}

To simulate the protocol under depolarizing noise, we select our values of $p$ and $n$ as before.
We then perform a density matrix ($\rho$) simulation of Simon's algorithm on a local device (i.e., the quantum communication channel is noiseless), applying a one-qubit depolarizing channel,
\begin{equation}
    \Delta_{1,p}(\rho) = \sum_{i=0}^3K_{i,p}\rho K_{i,p}^{\dagger},
\end{equation}
after each one-qubit gate and a two-qubit depolarizing channel,
\begin{equation}
    \Delta_{2,p}(\rho) = \sum_{i=0}^3\sum_{j=0}^3K_{i,j,p}\rho K_{i,j,p}^{\dagger},
\end{equation}
after each two-qubit gate.
Here,
\begin{equation}
    \begin{aligned}
        K_{0,p} &= \sqrt{1-\frac{3p}{4}}\sigma_0, \\
        K_{i,p} &= \sqrt{\frac{p}{4}}\sigma_i, \\
        K_{0,0,p} &= \sqrt{1 - \frac{15p}{16}}\sigma_0\otimes\sigma_0, \text{ and} \\
        K_{i,j,p} &= \sqrt{\frac{p}{16}}\sigma_i\otimes\sigma_j, \\
    \end{aligned}
\end{equation}
are Krauss operators and the $\sigma_{k\{k=0,1,2,3\}}$ are the Pauli matrices.
The results of this simulation are shown in Figure \ref{fig:depolarizing}.

Once again, success is certain when $p=0$, as the noise is not playing a role at all in the simulation.
That said, we see the probability of success decrease very quickly as a function of $p$ in this setting---it does not take much quantum noise to significantly impact performance.
This is likely due, at least in part, to correlations in the errors that emerge under the quantum noise.
These correlations are not possible under the classical noise models considered above.

\subsection{Discussion}

At this point, we have demonstrated a decrease in success probability as a function of both the noise parameter $p$ and the system size $n$ for each noise model.
As such, Guinevere can use this information to dictate how much noise is allowable in the network.
More specifically, assuming a given noise model, Guinevere can choose an oracle size $n$ to eliminate all devices operating with a noise parameter $p$ above a given threshold.

For example, consider the depolarizing model shown in Figure \ref{fig:depolarizing}.
By choosing an oracle of size $n=12$, Guinevere can assert that any client operating in a regime where $p>0.03$ will very likely be rejected from the network, and only clients operating with $p<0.01$ will likely be accepted into the network.
Clients with $p\in[0.01,0.03]$ have an appreciable probability of being accepted, and as such Guinevere should consider $p=0.03$ to be the effective threshold set by an oracle of this size.
Guinevere can decrease this threshold by increasing $n$ (or by requiring clients to pass multiple different oracles of size 12).

\section{Concluding Remarks}

In this paper, we have introduced a novel protocol based on Simon's algorithm for generating a verified network of quantum devices.
In other words, our protocol allows a quantum server to verify the presence of high-fidelity quantum information processing and communication on a remote client.
We demonstrated that a quantum server operating our protocol can successfully distinguish between a high-fidelity quantum device, a low-fidelity quantum device, and a classical device.
With this distinction made, the server can admit only the reliable quantum devices into the network.

This work can be extended in several ways.
First, it would be desirable to see an implementation of this protocol on a real quantum network.
This would certify the efficacy of this protocol when deployed on real physical systems more complex than the idealized noise models used to test the protocol in this paper.
This will also provide an opportunity to validate some of the implicit assumptions of this protocol, such as ready access to secure quantum communication channels.

Second, further work could be done on evaluating the protocol under more physically inspired noise models.
For example, one could construct a photon loss model which accounts for correlations in the losses of neighboring bits due to prolonged turbulence.
Additionally, one could account for turbulence on both the incoming and outgoing quantum messages.

Finally, in this paper we list a few variations that could be made to the basic server-client model of this protocol.
Certainly additional variations are possible, and some of them may have advantages over the basic model we have described here.
Thought should be given to exploring other variations of this work, and particular emphasis should be given to directions inspired by classical state-of-the-art secure classical network communications.

\acknowledgments{
    S.D. acknowledges support from the John Templeton Foundation under Grant No. 63626.
    R.R. thanks Agung Budiyono for the impetus to ask the essential question behind this work: ``How can Simon's oracle be both unknown and implementable?''
    All authors thank Thomas Barthel for thoughtful questions and helpful comments that improved this paper.
}

\section*{Data Availability Statement}

All code and data used to produce this paper are available at \url{https://github.com/patsej07-afk/Quantum-Networked-Simons}.

\section*{AI Usage Statement}

The authors did not use generative AI in this paper.

\appendix

\section{Approximate Protocol}\label{sec:approximate-protocol}

\begin{figure}[b]
    \centering
    \includegraphics[width=\linewidth]{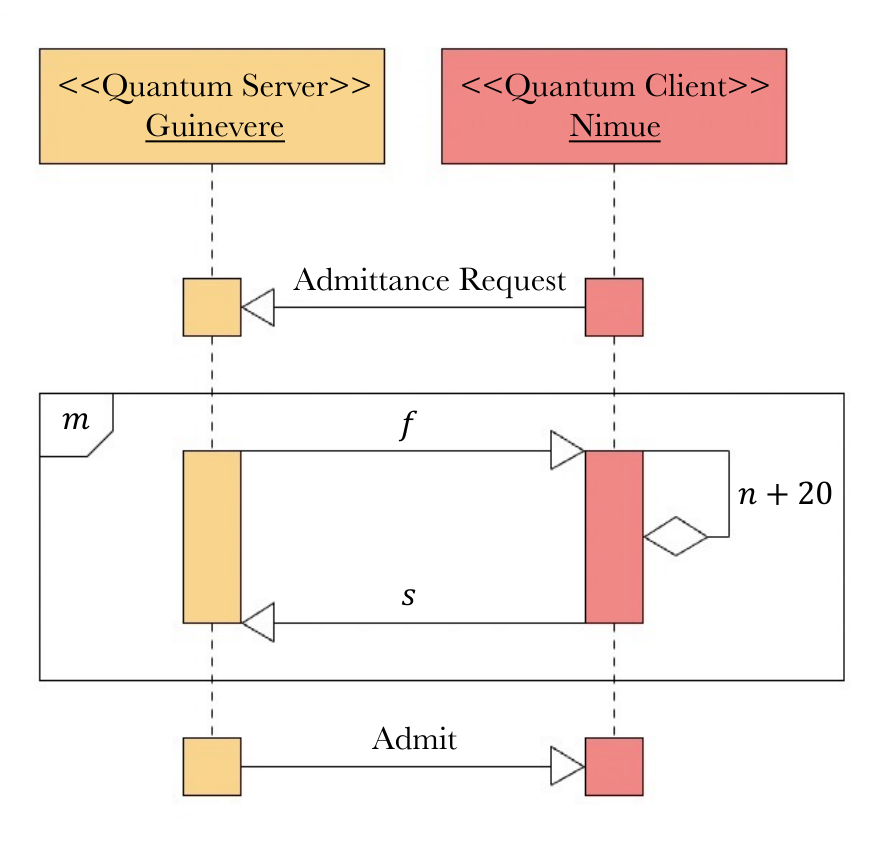}
    \caption{
        \textbf{Classical communication protocol.}
        An approximation of the network verification protocol using only classical communication, denoted by triangular arrows.
        Nimue performs $n+20$ local evaluations of the quantum oracle, denoted by a diamond arrow.
        In lieu of the capacity to track the number of oracle queries, Guinevere admits Nimue only if she returns the proper $s$ within a specified number of minutes, $m$.
    }
    \label{fig:approximate-protocol}
\end{figure}

The protocol in Section \ref{sec:methods} assumes physical infrastructure allowing direct communication of quantum information.
The components of this technology have been demonstrated experimentally, however, a widespread implementation of this technology does not exist.
As such, this section describes an alternate protocol, outlined in Figure \ref{fig:approximate-protocol}, requiring only classical communication.

The protocol begins when Nimue (or Lancelot) signals to Guinevere a desire to join the network.
First, Guinevere selects a problem size $n$, and then instantiates a Simon oracle $f$ with a hidden string $s$.
Guinevere also selects a time limit, $m$, for the protocol.
Since no quantum information can be communicated, Guinevere then sends the oracle $f$ to Nimue over a classical channel.
Once she has received the oracle, Nimue implements it on her quantum computer and queries it $n+20$ times to construct a system of equations which she can solve to recover $s$.
This done, Nimue returns her reconstructed $s$ to Guinevere.

When Guinevere receives the reconstructed $s$ from Nimue, she evaluates it to ensure its correctness.
Guinevere also tracks the elapsed time between her message to Nimue and Nimue's response.
Since Guinevere has no direct control of the number of queries Nimue makes to the oracle, adherence to the time limit is the best proxy Guinevere has for the number of oracle queries.
If both $s$ is correct and the elapsed time before Nimue's response is below $m$, Guinevere will admit Nimue into the network.
If either of these conditions is not true, then Guinevere will reject Nimue (or rather Merlin or Morgan) from the network.

Now, this protocol has one major weakness---it requires Guinevere to share the details of the implementation of the Simon oracle with the client.
As soon as the information is shared, Guinevere has no control over what the client does with it.
In other words, there is nothing stopping Merlin from using his classical resources to reverse engineer the oracle without querying it at all; something that he is able to do because he knows exactly how the oracle is implemented.
Moreover, if Merlin possesses sufficiently fast classical resources, he may be able to complete this process and return $s$ to Guinevere within the appropriate time limit $m$.
Therefore, this modified network protocol cannot verify quantum network integrity.
Rather, its utility is as a pedagogical tool for building intuition about the secure protocol.
Moreover, an implementation of this protocol built for execution on IBM devices and simulators is provided in the project GitHub repository.

\bibliography{bib}

\end{document}